\documentclass{sig-alternate}
\newcommand{\eqdef}{\ensuremath{\stackrel{\mbox{\upshape\tiny def}}{=}}}
\newcommand{\arrowNamed}[1]{\ensuremath{\stackrel{#1}{\rightarrow}}}

\begin{document}
  \def\style{newapa}   
\conferenceinfo{TBA}{}
\CopyrightYear{2022} 

\title{The Logic of Collective Action Revisited
}
%
%

%

\author{
%
\alignauthor {Ian Benson} 
\\
       \affaddr{School of Education}\\
       \affaddr{Froebel College}\\
       \affaddr{University of Roehampton}\\
       \affaddr{London}\\
       \email{Ian.Benson@Roehampton.ac.uk }
}
\maketitle
\begin{abstract}
Mancur Olson's ``Logic of Collective Action'' predicts that voluntary action for the common good will only happen in small groups. His theory of the structure and behaviour of organisations fails to account for the UK Labour Party's promotion of, rejection and ultimate compliance with its Political Parties Act (PPERA). I propose a revised computational theory to explain this behaviour. It negates key tenets of Olson's work: that consumption of a common good by one member inevitably reduces the quantity available to another and that negotiation between members does not in general affect a decision to work for the common good. The approach has application in private and public sector service design.

\end{abstract}



\section{Introduction}
\label{sec:one}
In his seminal work ``The logic of collective action: public goods and the theory of groups'' economist and political scientist Mancur Olson suggests that there are essentially two, mutually exclusive, forms of collective action - voluntary or coerced. His theory predicts that voluntary action for the common good will only arise in small groups and then only in certain circumstances. Coercion is necessary in almost every case because each organisation produces goods (relationships) that are commonly available to all their members. So ``a state is first of all an organisation that provides public goods for its members, the citizens'' and ``other types of organisations similarly provide collective goods for their members''~\cite{b1}. If organisations fail to provide these benefits then they will eventually die. Since common goods are by definition available without constraint to qualified individuals (citizens, employees) there is no incentive for any one individual to participate in the work of the organisation voluntarily.

We revisit Olson's key assumptions on the behaviour of people in organisations. These are:

\begin{enumerate}
\item that consumption of a common good by one member reduces the quantity available to another and
\item that negotiation between members does not in general affect a decision to work for the common good
\end{enumerate}

I propose an alternative computational theory and illustrate this through a study of membership activity to record and report political donations of volunteer effort, cash or kind (see  \texttt{http://www.electoralcommission.org.uk} ).

\subsection{Background}

The public-private relationship is increasingly mediated by information systems. Such systems require validation -- that is, the state needs to know that stakeholders are satisfied that the system meets their requirements, and verification -- the public needs assurance that they are implemented as secure and trustworthy software. Information flow security is a subject of active research, including language-based security and formal methods \cite{b1o}. Verification of large or unbounded concurrent and distributed systems has been a widely studied for at least three decades. It has  impressive existing literature \cite{b1p}. In this paper we focus our attention on automata-based validation, where a system is described by a state space, or labeled transition system, agreed with the sponsor, and correctness is expressed in terms of relationships between automata. Verification follows by deriving of a set of test scenarios from the paths in the transition system and applying these to the constructed information system.

The question of the nature of the relationship between state and civil society is a subject of active research. Banushi et al identify the simplification of application procedures for national schemes to build trust and minimise the time and cost of application as being essential to improve the performance of an initiative to make grants and credit guarantees to farmers \cite{b1a}.  In a recent paper on public crisis management Wang analyses the response of the Chinese state and voluntary organisations to the Covid-19 pandemic. She argues for a `cobweb' rather than an `inverted T' operation mode for public-private collaboration. Her analysis draws on Coase and Williamson's institutional economic models for the firm, markets and hierarchies. In this framework the firm and the state are optimised to efficiently conduct business where there are clear tasks, well defined work and high transaction frequencies. In the face of unconventional circumstances informal organisations have an advantage because their flat and flexible organisational structure can be readily adjusted \cite{b1b}.

Benson et al~\cite{b6} and Ciborra~\cite{b7} combine institutional economic analysis with software engineering methods.  The components of this approach are elaborated in the bibiography. We take as our basic transaction model Winograd and Flores' Conversation for Action \cite{b2}. Our perspective on the decline of volunteer activity and the conditions for its revival is taken from Putnam \cite{b3} and Resnick \cite{b4}. Weigand~\cite{b8} reports on the coming of age of Language/Action theory. Milner~\cite{b9} illustrates how to model and reason about computation as communication with concurrency, and Pavlovic ~\cite{b10} describes how to exploit the duality of operational semantics and logic in a software engineering practice that uses tests (we call them \textbf{decisions}) as specifications.

Stephen Uttley, Labour's Director of Finance and statutory Party Funding Manager, said the system ``gives the Labour Party confidence that captured donations will be reported in compliance with the PPERA, 2000" \cite{b5}.

\subsection{Structure}

\begin{itemize}
\item In Section~\ref{two}, we recall the vocabulary of organisational objects, operations and relationships that Olson developed to reason about membership associations. These represent why an organisation exists (we call it \textbf{an accounting unit}) and those who are involved (\textbf{donors, etc}). We recount the criteria people adopt when they decide whether to work for a common good, together with Olson's group cost/benefit analysis of the circumstances in which they may choose to participate or abstain. Olson rejects negotiation as a factor that can affect organisational behaviour. We reproduce his argument by reference to Winograd and Flores' \textbf{Conversation for Action}\cite{b2}. This is a `choreographed dance' between two people witnessed by a third in which bilateral speech acts set out time, cost and quality conditions of satisfaction for a task.
\item Then, in section~\ref{three} we follow Olson's generalisation from a single unit to a group of units and apply the resulting analysis to the work of the Labour Party activists.  Labour exhibits many of the characteristics predicted by Olson and observed by Putnam \cite{b3} and Resnick \cite{b4}, such as declining levels of membership and membership activity. In our study the common good is the need for all units to comply with the Political Parties, Elections and Referendums' Act (PPERA, 2000). According to Olson since the Party is not a small organisation, its units should either conform to PPERA through coercion or additional legal sanctions are needed. We describe the intention of PPERA and its sanctions, analyse the work that each Party unit needs to do as a generalisation of a Conversation for Action (\textbf{CfA}), and reproduce an account of their behavior from the UK national press. We observe a pattern of reluctant compliance that cannot be explained within Olson's framework, and conclude that his assumptions about the nature of decision making and the common good need to be revised.
\item 
To explain what happened we need to look again at Olson's account of negotiation within and between units. In Section~\ref{subsect:decide} we introduce a correct by construction composition of CfA to form a collective \textbf{decision protocol} -- a quarterly cycle of inter-unit work. The consent of initially reluctant Party units was ultimately won by building a system for recording and reporting that gives confidence that all recorded donations are properly reported ~\cite{b5}. Because the recording process highlighted those units that were not correctly reporting donations it induced them to fall into line.
\end{itemize}
 
We found that accountability was ultimately achieved through negotiation that built confidence in the correctness of the decision process. Confidence is a category of common good essential to the operation of teams, markets and systems. Furthermore, one unit's confidence is not gained at the expense of another. This negates Olson's key tenets  --  that organisational units only enjoy common goods at each other's expense and that negotiation does not affect behaviour.

%

\section{Olson' s Logic of Collective Action}
\label{two}
\begin{figure}[ht!]
\centering
\leavevmode
\includegraphics[width=0.35\textwidth,origin=c]{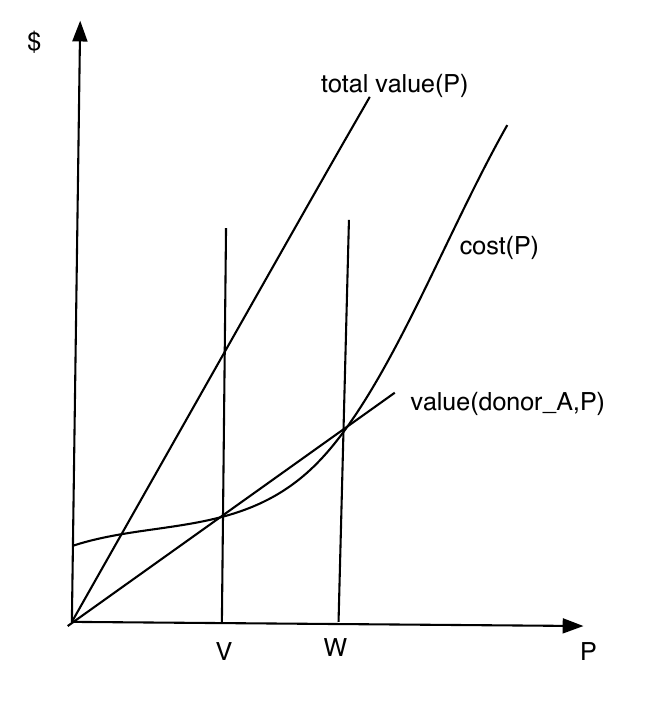}
\caption{\it Olson's group cost/benefit analysis (Value vs Production level)} 
\label{y}
\end{figure}

Mansur Olson's analysis of the conditions for successful collective action for a common good is shown in Figure~\ref{y} \cite[p 32]{b1}. 

He considers an organisational accounting unit \textbf{AU}, the people involved (such as $donor_A , donor_B$) and the decisions $\delta(donor_A,P)$,\;$\delta(donor_B,P)$ that A, B make to participate at some level of production \textbf{P}. $\delta(\_,\_)$ is a predicate, a logical function that returns \texttt{true} or \texttt{false}.

The graph shows his group cost/benefit analysis. The y axis is cash  (\$) and the x axis the level of production (P). Olson plots the common good -- that is the \textit{total value} equal to $ \sum\limits_{donor}{value(donor, P)} $ -- as a straight line and the cost of production \textit{cost(P)} as a J shaped graph, with the initial fixed cost of the first unit  \textit{cost(0)} increasing with the volume P produced. 

\subsection{The structure of negotiation in small groups}

\begin{figure}[ht!]
\centering
\leavevmode
\includegraphics[width=.5\textwidth,origin=c]{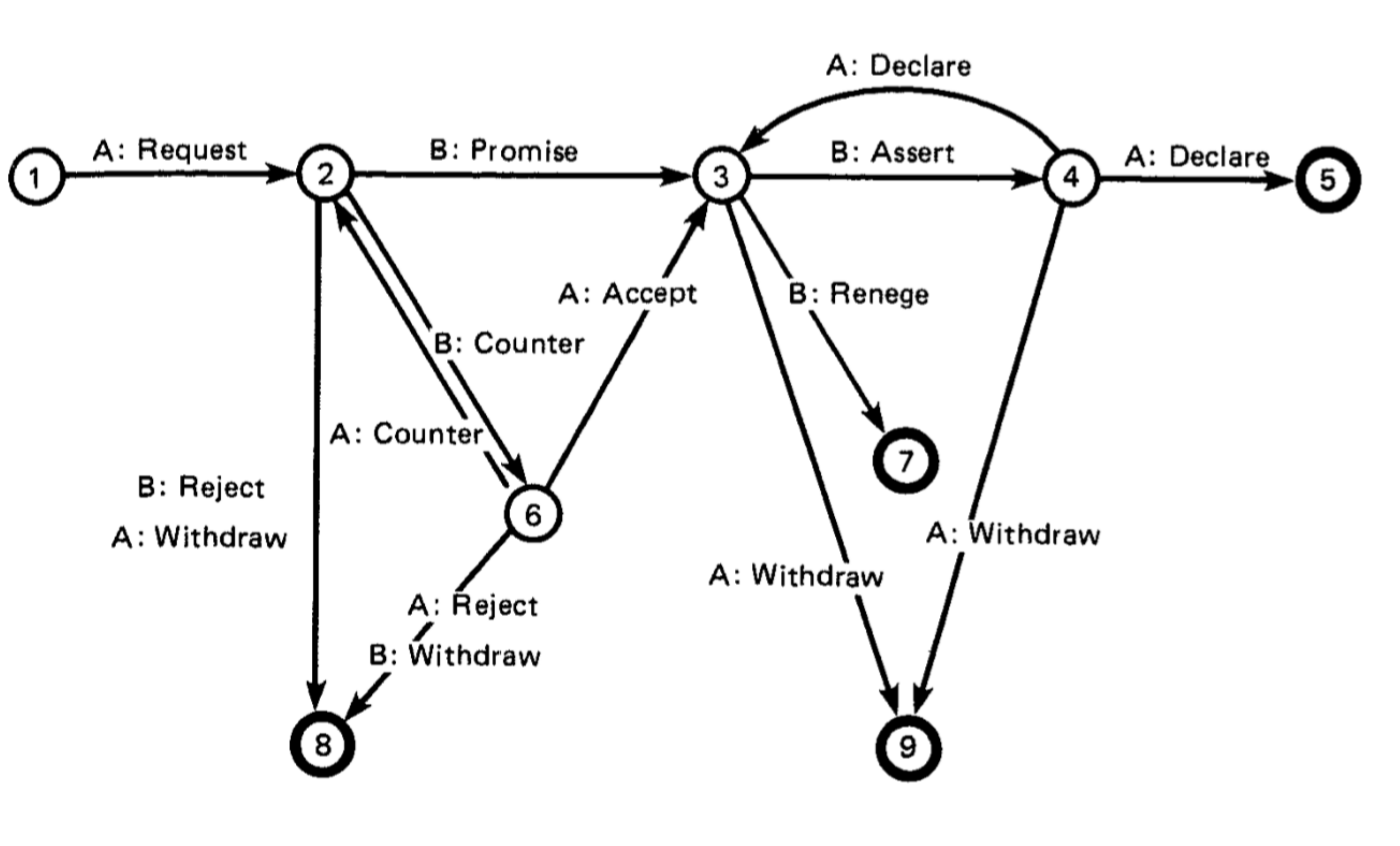}
\caption{\it Conversation for Action}
\label{CfA}
\end{figure}

Olson considers two cases: a $donor_A$ whose share of the total production value exceeds \textit{cost(P)} at a level of production in the range [V,W], and a $donor_B$ whose share is never greater than \textit{cost(P)}. We take Winograd and Flores language/action theory as a model for negotiation and bargaining in small groups. They draw on Searle's speech acts to classify utterances into a small number of categories which fall into a particular pattern in discourse. The structure of a typical Conversation for Action is shown in Figure~\ref{CfA}, a state change protocol in which arcs are labelled by person (A or B) and type of speech act. 

Olson assumes that there are only two classes of donor. He dismisses the possibility of  ``strategic bargaining''  which might induce two donors to collaborate to meet the \textit{cost(P)}. He argues that the incentive for anyone to actively participate in creating the common good only exists in the circumstance of $donor_A$ and then only in the interval [V,W]. He says that bargaining will not be a viable alternative as  $donor_A$ and  $donor_B$ will inevitably differ in the proportion of the common good that they will gain and the net benefit of meeting the cost of this negotiation. Consider the case in which A gains much more than B from the common good. All A can do to get B to comply is to threaten the smaller member saying, in effect, if you do not provide more of the common good I will provide less myself, and you will be worse off as a result. This is an empty threat since \texttt{B:Counters} that A will suffer more from the reduction in production since A has more to gain. Another factor that makes the threat less credible is that the maximum amount of the common good than B can contribute is small, and may ourweigh A's cost of bargaining. This is the opposite case for B who has an incentive to prolong the negotiation, hence Olson predicts either \texttt{B:Rejects} or \texttt{A:Withdraws}~\cite[p 12]{b1}.

One consequence of Olson's analysis is that the size of a group will influence the likelihood that it will provide its members with a collective good without coercion (apart from access to the common good itself). In general it is only sensible for individuals to engage voluntarily in group activity at certain stages in the inception phase of small groups when the cost of production can be met by one member alone. Once a group becomes established then each new member will benefit from the common good without active engagement. He argues that this is why people may vote but do not join political parties (unless they seek office themselves), why trade unions need the coercion of the ``closed shop'' and why the state needs laws of taxation to survive. His analysis provides a plausible explanation for Putnam's observations on the decline of mass associations~\cite{b3}.


\section{``Labour Chief says he fears jail over sleaze''}
\label{three}
In this section we apply Olson's analysis to the group of accounting units that do the work of the UK Labour Party. At first glance the Labour Party exhibits many of the characteristics predicted by Olson and observed by Putnam -- from decreasing levels of membership to declining membership activity. However we find on closer inspection that Olson's theory of groups cannot account for the observed pattern of behaviour (section~\ref{sec:foura}). This leads us to question his assumptions about the nature of the common good and the significance of negotiation. 

\subsection{Olson's theory of groups}

Olson recognises that any organisation or group will usually be divided into subgroups or factions that may be opposed to one another. Nevertheless he argues that this does not weaken his assumption that organisations exist to serve an overarching common interest of their members. ``The assumption (of common interest) does not imply that intragroup conflict is neglected. The opposing groups in an organisation ordinarily have some interest in common (if not why would they maintain the organisation?), and the members of any subgroup or faction also have a common interest of their own'' ~\cite[p 13]{b1}.

He adapts his framework to account for the conflict within groups and organisations by considering ``each organisation as a unit only to the extent that it does in fact attempt to serve a common interest, and considers the various subgroups as the relevant units with common interests to analyze the factional strife'' ~\cite[p 13]{b1}.

%

In our study the common good is a need for every accounting unit to comply with the Political Parties, Elections and Referendum's Act (PPERA, 2000) passed into law by the 1997 Labour Government. This act was a political response to a series of Tory funding scandals. The law came into effect in February 2001. We review the goal of PPERA, its sanctions and the work that needs to be done to ensure compliance. 

The objective of the law was to record and aggregate donations received from a donor at any of approximately 700 accounting units every calendar quarter and report if they total more than a certain amount. Responsible party officers, a new statutory role, could be fined up to \pounds 5,000 on each count of non compliance and jailed for a year (although imprisonment was considered unlikely). The circumstances in which donations are reported depend on which unit receives the donation, how much the donor has given that year, and when it was accepted.

\begin{figure}[ht!]
\centering
\leavevmode
\includegraphics[width=.5\textwidth,origin=c]{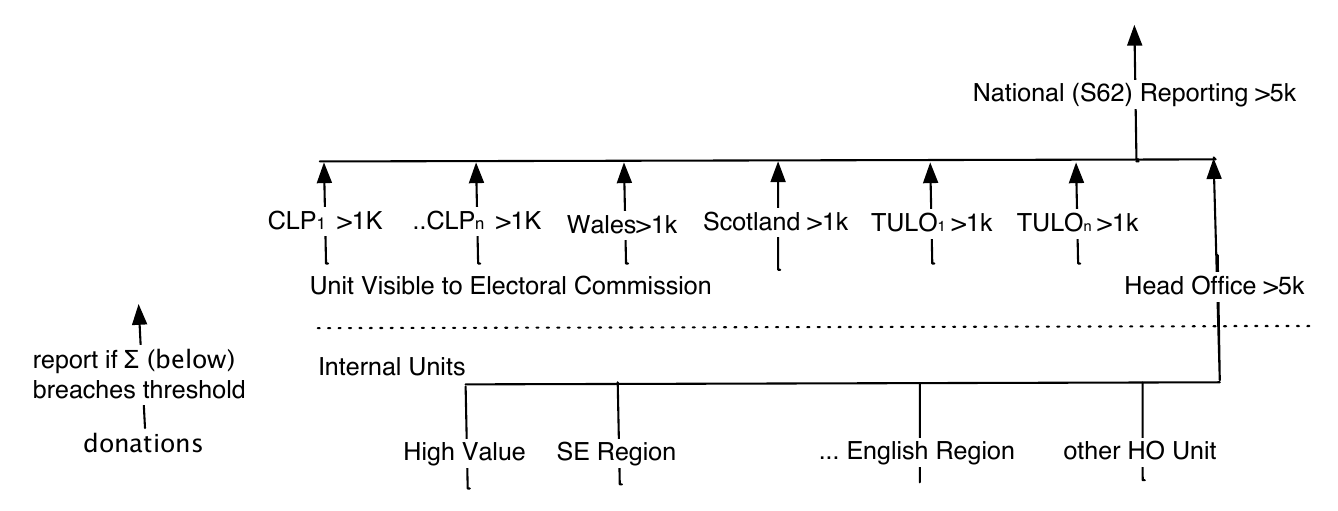}
\caption{\it Composition of Quarterly Decision Protocols}
\label{0}
\end{figure}

In order to comply with the Act the Party superimposed statutory officers and accounting units on its constitutional structure of Constituency Labour Parties (CLP), trade union liaison organisations (TULO), national parties (Wales, Scotland) and English Regions. These new statutory posts are called Responsible Officers. The statutory reporting structure, and the relevant thresholds for reporting each quarter on donations received at a accounting unit from a donor are shown in Figure~\ref{0}. This shows the parallel and sequential nature of decision making that the Party needs to perform.

\subsection{Task Analysis: Roles and Decisions}
\label{sec:three}

Every calendar quarter the Party Funding Manager takes four steps to prepare the Party's Report to the Electoral Commission. These decision rules are in compliance with Statute and Regulations and the practice that has developed since 2001. They are:

\begin{enumerate}
\item 
For the Head Office accounting unit, for each donor and donation, test if the amount donated is \textit{recordable} for reporting (that is more than \pounds 200), if not
\item 
for each local accounting unit, for each donor and donation, test if the amount donated is recordable for reporting, if not
\item 
for each donor and donation aggregate the total amount donated this year to date and test if the aggregate exceeds the National (also known as the \textbf{virtual AU} or Section 62(12)) threshold. If so report in aggregate recordable donations (ie those not reported in steps (1) or (2) or in prior quarters) to the Electoral Commission, and the constituent donations thereof in an internal \textbf{Section 62 audit report}, and
\item 
finally, record by donor and accounting unit those donations that are not recordable in an internal \textbf{Carried Forward report}
\end{enumerate}

\subsection{Test Paths}
\label{sec:foura}

\begin{figure}[ht!]
\centering
\leavevmode
\includegraphics[width=0.5\textwidth,origin=c]{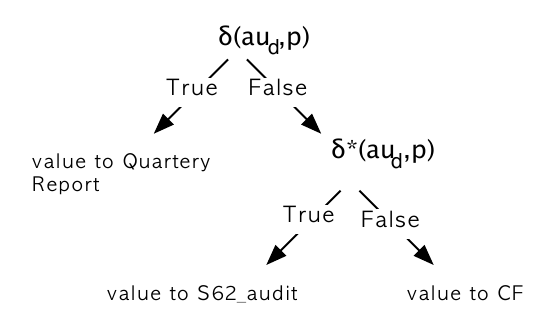}
\caption{\it Individual Donation Decision Tree (period p, donor d, unit au)}
\label{ii}
\end{figure}

The Quarterly decision tree in  Figure~\ref{ii} shows how we can encode the reporting state of a donor to an accounting unit in a period using just two predicates. When interpreted in the context of test data, predicates are a logical function of accrued data. Since there are 3 possible outcomes each period, we need as a minimum two predicates to encode the donation reporting state for each donor, unit and quarter.

A donor's report with respect to a physical accounting unit consists of zero or more quarters in which all the recordable donationså are carried forward (\textbf{c}), followed by zero of more quarters of Section 62(12) reporting (\textbf{s}), followed by zero or more Quarterly Reports that show their donations as breaching the threshold of a physical accounting unit (AU) (\textbf{r}). We can encode these constraints as a permissible path by adopting Party Funding Manager Stephen Uttley's speech act notation: c for a quarter in which reporting is carried forward, s for a Section 62(12) report and r for a regular quarterly report. 

\begin{figure}[ht!]
\centering
\leavevmode
\includegraphics[width=0.5\textwidth,origin=c]{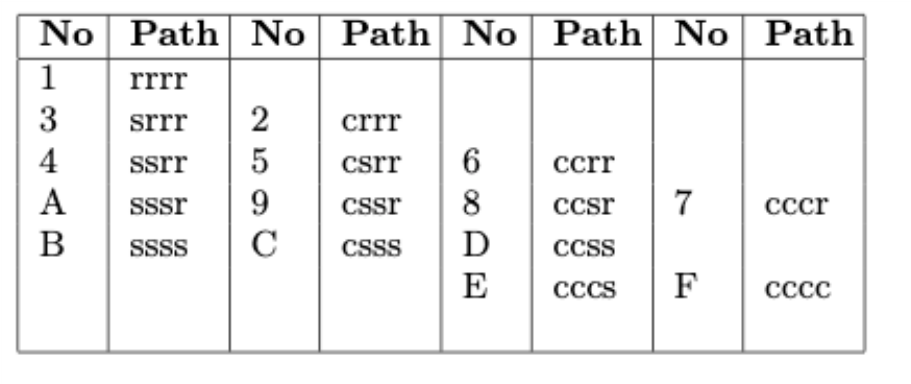}
\caption{\em AUdonor Paths through Reporting Space}
\label{objectives:summary}
\end{figure}

A \textbf{path name} is a permissible path string, consisting of four letters, corresponding to the sequence of transitions taken to reach a terminal state (Figure~\ref{iii}). A path name is built up by appending the speech act taken each quarter to the end of the ``path string'' eg, c, cs, csr, csrr. We can conveniently abbreviate these 15 path names using the hexadecimal numerals 1 to F as shown in Figure~\ref{objectives:summary}.

%

\subsection{Observed Behaviour}
\label{sec:foura}

The Funding Manager's decision logic relies on correct recording. It is a matter of public record that this did not happen in the first two annual cycles of reporting.

%

The Guardian newspaper reported on 13th February 2003 that ``All the political parties have been given a warning by Sam Younger, the chairman of the Electoral Commission, that they must comply with regulations and register donations in time or face prosecution. The warning comes after Labour failed to report \pounds 236,952 on time for the first quarter of 2002, and \pounds 1,815,549 for the second quarter. The sums were eventually reported in the third quarter. Another \pounds 20,192 which Labour should have been reported previously was included in the returns for the fourth quarter. The Conservatives also found themselves in trouble with the commission after refusing to disclose the name of their new biggest donor.'' ~\cite{b11}

Under the headline ``Labour chief says he fears jail over sleaze'' the Kevin Maguire~\cite{b12} reported:

\begin{quotation}
Labour's general secretary has warned he could be heavily fined or even
jailed over a growing number of breaches by the party of an anti-sleaze
law Tony Blair introduced to clean up political donations.

David Triesman admits in an internal report, marked ``private and
confidential" and seen by the Guardian, that Labour's electoral
prospects will be harmed if the full extent of the crisis is made
public.

Boasting that it has been largely kept under wraps except in Scotland,
he says the ``problems are getting worse'' with more Constituency Labour
Parties failing to file accounts or declare donations worth \pounds1,000 or
more.

Mr Triesman describes the position as ``serious and intolerable'' and
accepts that it is embarrassing as well as illegal when the party fails
to comply with a law introduced by Labour after a series of Tory funding
scandals.

``Politically it is portrayed as party illegality in relation to our own
legislation.

``We have managed media effectively except to any great degree in
Scotland where it is simply hostile and will now have an impact on
electoral prospects,'' says the Triesman report.

``In strict legal terms, matters are still more serious. First, we are
already at risk of significant fines. Second, I am increasingly at risk
of criminal action and I assess this risk as higher if no credible steps
are taken to do a huge amount more than at present to demonstrate that I
am not willingly or recklessly failing to intervene.''

An investigation by Labour's Old Queen Street HQ into the finances of
local Labour parties during the 18 months to September 2002 is
understood to have uncovered more than 100 potential breaches of the
law.

Donations worth more than \pounds1,000 must be declared but the inquiry
discovered an alarming number of constituencies either did not file
accounts or lodged inaccurate returns.

The general secretary is to propose at this month's Labour spring
conference in Glasgow that a single party bank account be introduced for
constituencies to allow officials to offer advice to voluntary
treasurers and ensure accounts are filed correctly.

Mr Triesman wrote a letter of apology last month to Sam Younger,
Chairman of the Electoral Commission set up under the Labour legislation
to regulate donations, and the Whitehall-appointed body is due to report
next week.
\end{quotation}

\subsection{Transition System Semantics for Workloops}
\label{sec:five}
To encourage correct recording the Funding Manager sponsored the creation of a system for computer-mediated communication to overcome the breakdowns in recording that had put the General Secretary at risk. To develop a test suite specification the structure and behaviour of the Party was storyboarded with two commonplace diagrams. The structure of decision making was drawn with the conventions of the London Underground: train lines represent phases of a Conversation for Collective Action (these are the same phases as those in the CfA), interchange stations represent decisions, and data is represented as coloured tokens. Coloured ``fare zones'' represent the roles of Responsible Officer and Funding Manager. Decision making at each interchange station is drawn as an ``act-decide transition system.'' That is, a transition system labeled by speech acts, whose states are marked by formulae true in the prior state. The organisation of the Party as a whole is then taken to be the sequential composition of the four quarterly decision protocols that make up the annual reporting cycle.

\subsubsection{Decisions are Protocols too}
Winograd and Flores' general Conversation for Action model for communication centres on the idea of a negotiated time, cost and quality ``condition of satisfaction.'' This is the basis for the initial \texttt{A:Request} act, and the grounds for A to ultimately \texttt{declare} that the action is complete. In the case of PPERA the Condition of Satisfaction is that``for this pair of donor and receiving Accounting Unit, on the last day of each calendar quarter, either make a regular report to the Electoral Commission, a Section62(12) report or carry the recorded donations forward.'' But this model does not record what rules will be used to decide which report to create, or how the conversations will interact with each other to determine whether Section62 (National) reporting will come into play.

\begin{figure}[ht!]
\centering
\leavevmode
\includegraphics[width=.5\textwidth,origin=c]{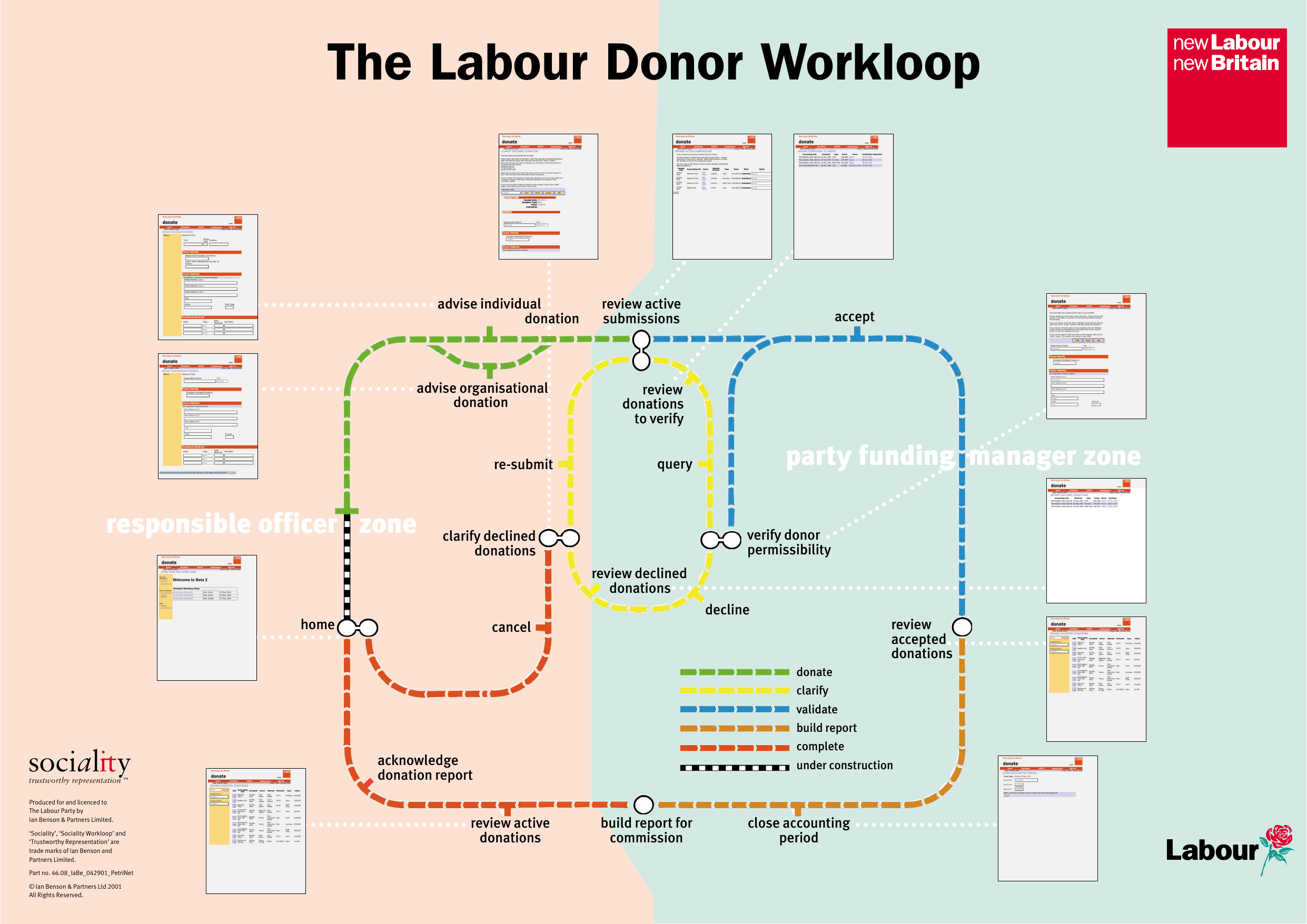}
\caption{\it Labour donor workloop: a storyboard for c2800 labour donor decision protocols}
\label{xx}
\end{figure}

\begin{figure}[ht!]
\centering
\leavevmode
\includegraphics[width=0.5\textwidth,origin=c]{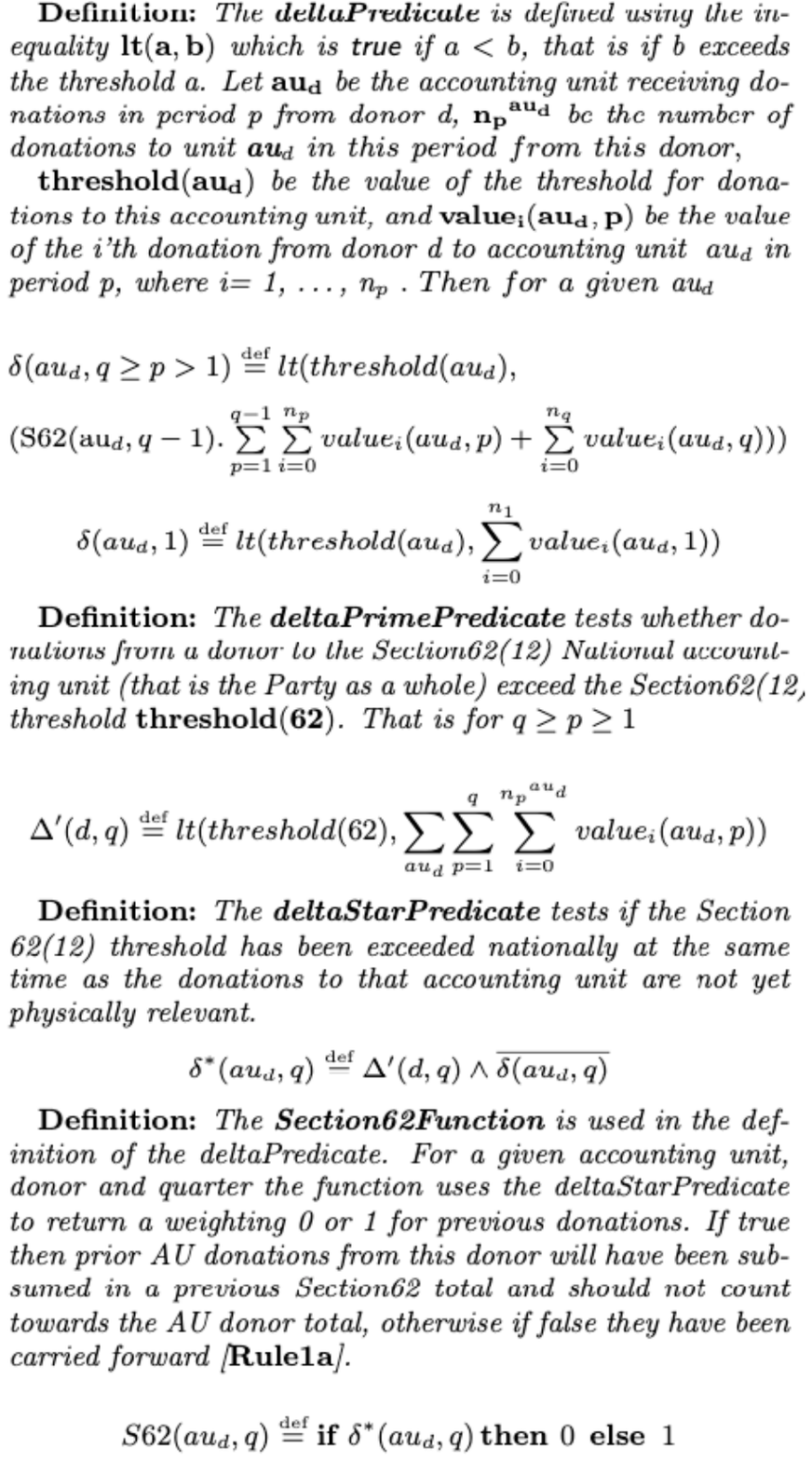}
\caption{\it Definitions for Decision Logic}
\label{iii}
\end{figure}

We call the parallel composition of individual Conversations for Action between donors and accounting units which resolves these questions the Labour Donor Workloop (Figure~\ref{xx}). In the workloop communication is by handshake, with no latency, just like the transitions in the CfA state change conversation. We model the decisions by A and B on whether and how to proceed by a decision agent located at each CfA state. This agent is illustrated in Figure~\ref{iii} -- a labelled transition system for the workloop interchange station Build Report for Commission in Quarter 4. The Figure shows the 15 permitted paths for an accounting unit's report. Let $ \alpha $ be the speech act: r, s or c. Then we write ${\arrowNamed{\alpha}  \textit{$\delta_p$}}$ for the p'th decision $\delta_p$ that the action $\alpha $ be taken. As in Figure~\ref{ii} states are labelled by formula over two predicates  {$ \delta_p $} and {$ \delta^*_p $}. The start of the year is the leftmost node in the graph. Transition arcs are labelled by the action taken with the donor's donations to the AU when the quarter is closed. Unlabelled states take the formula \[ \lnot (\delta_p \vee \delta^*_p) \]. We can infer from this that the decision made in the prior state was to make a carried forward report.

As the year progresses there is a need to process retrospectively donations that come to light after the end of a quarter. This means that the individual quarter decision agents become increasingly complicated. (Figure~\ref{iv} to \ref{vi})

\begin{figure}[ht!]
\centering
\leavevmode
\includegraphics[width=0.5\textwidth,origin=c]{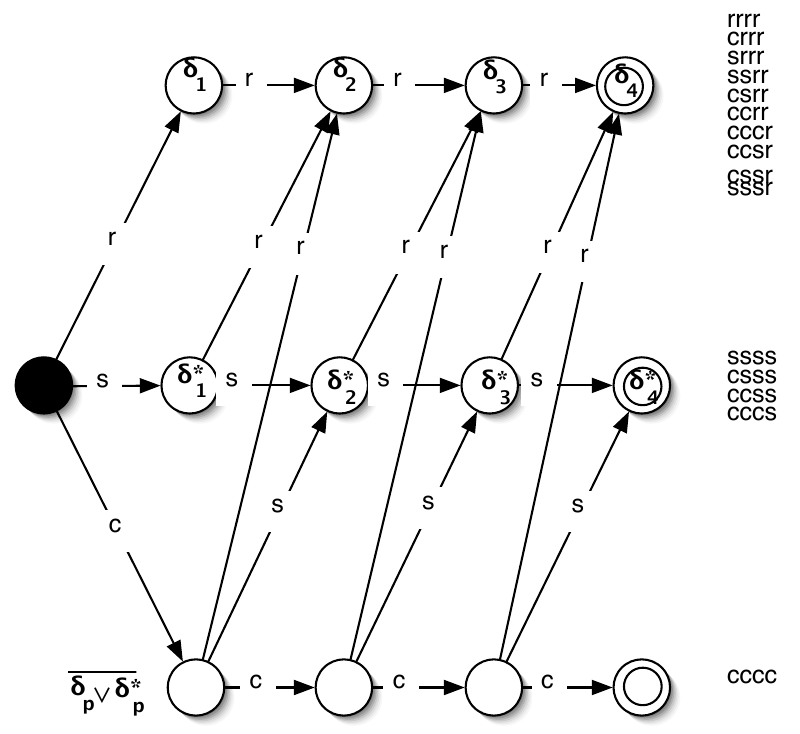}
\caption{\it AUDonor act-decide labelled transition system (Q4)}
\label{iiii}
\end{figure}


\section{Decision Protocols}
\label{sec:six}

We rely on the combination of our set of decision agents and their tests to prove that decision making covers all eventualities. We cannot rely on observation alone to demonstrate this, since for each donor we have to consider the interaction between the virtual AU and perhaps 700 physical AU aggregation levels (Head Office thresholds and local AU donor thresholds). 

To show that the decision agents together do indeed satisfy the goal of the legislation we introduce a number of formal definitions in \textbf{bold} font (section~\ref{subsect:decide}). These correspond to custom, practice and Statute. The key Section 62(12) of PPERA which deals with National donations is reproduced at http://sociality.tv/s62.

Our task is simplified by the fact that the set of donations from a donor to an AU in a given period are all treated in the same way. We can therefore reason about the sum of these donations as if it was a single donation without compromising the integrity of a formal proof. Secondly we can assume that the Head Office units - although independent as receiving units - are part of a single accounting unit with a single threshold. We can therefore consider all donations from a donor to any Head Office unit as a single compound donation to Head Office itself.

Furthermore, the interaction between local, Head Office and Section 62 is carefully constrained. There is no connection between reports of a donor's activity with two different physical AU except the coupling demanded by Section 62(12). 

To prove that this reasoning is correct we further refine the storyboard workloop to form a quarterly \textbf{decision protocol}. This consists of the following elements:
\begin{enumerate}
\item
The set of all recordable donations of cash or kind from a donor in a period to an AU gives rise to two concurrent Conversations for Action: one between the donor and the receiving accounting unit and one between the donor and the national (virtual) accounting unit. We call this compound CfA, along with its associated data and decision logic, a  \textbf{Conversation for Collective Action (CfCA)}.
\item
Each CfCA decision is modelled as a first order predicate calculus formula. This tests for an expected reporting outcome.  
\item
For each CfCA decision agent we develop a suite of test cases to exercise all legal recording and reporting outcomes in the quarter. We prove that there are no other test cases.
\item
For the CfCA as a whole there is a proof of test coverage. That is, we prove that the decision logic (1) considered as a logical theory satisfies the test formulae (2) in each scenario (3).
\end{enumerate}

\subsection{Decision Logic}
\label{subsect:decide}

We illustrate our approach by constructing decision agents for reporting to the Electoral Commission at the end of the first and fourth quarters. To implement the rules for physical AU reporting (section~\ref{sec:three} Rules (1 and 2)) we construct the deltaPredicate {$ {\bf \delta(au_d,q)} $} over the set of donors {$ \bf {d} $}, units \texttt {$ \bf {au_d} $}  and  reporting quarters {$ \bf \texttt {q, p} = 1, \ldots , 4 $}. Informally, if {$ {\delta} $} is \textsf{true} then this signals that the donations received in the unit from the donor will be reported this quarter.

Section 62(12) and carry forward reporting (Rules 3 and 4) use the deltaStarPredicate {$ {\delta^*(au_d,q)} $}. This is \textsf{true} if the cumulative total to date of all the donations received at any unit exceeds the Section62(12) (National) threshold of \pounds 5k and the unit {$ {au_d} $} is not reporting. 

The deltaPrimePredicate establishes whether the cumulative total to date of all the donations received at any unit exceeds the Section62(12) (National) threshold of \pounds 5k. It is used in the definition of the deltaStarPredicate.

The Section62Function {$ {S62(au_d,q)} $} returns zero or one, depending whether a donor is to be reported as Section 62(12) for their donations to this unit this quarter (0), or is to report via the AU, or have these donations carried forward (1). It uses the deltaStarPredicate and is used in the formal definition of the deltaPredicate to exclude donations reported as Section62(12) from the physical AU threshold test.

We write {$ \delta $}, {$ {\delta^*} $} and {$ {\Delta^\prime} $} for these predicates to signify circumstances in which they are \textsf{true}. {$ \overline \delta_p$} {$ ( \overline \delta^*_p $}, {$ \overline \Delta^\prime_p )$} signify a quarter, p, in which the {$ \delta_p $}  {$ ({\delta^*_p} $} ,   {$ \Delta^\prime_p ) $} predicate is \textsf{false}.

\subsection{Test Formulae}
\begin{figure}[ht!]
\centering
\leavevmode
\includegraphics[width=0.5\textwidth,origin=c]{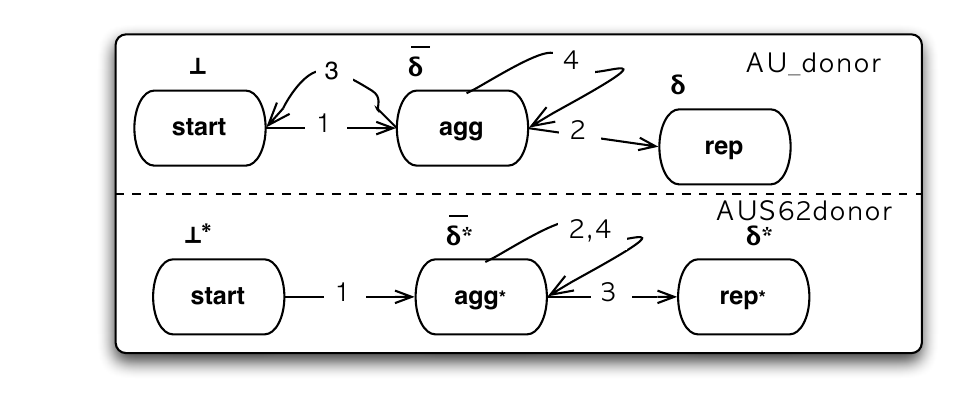}
\caption{\it Individual Donation Reporting State Chart (Q1)}
\label{iv}
\end{figure}

\begin{figure}[ht!]
\centering
\leavevmode
\includegraphics[width=0.5\textwidth,origin=c]{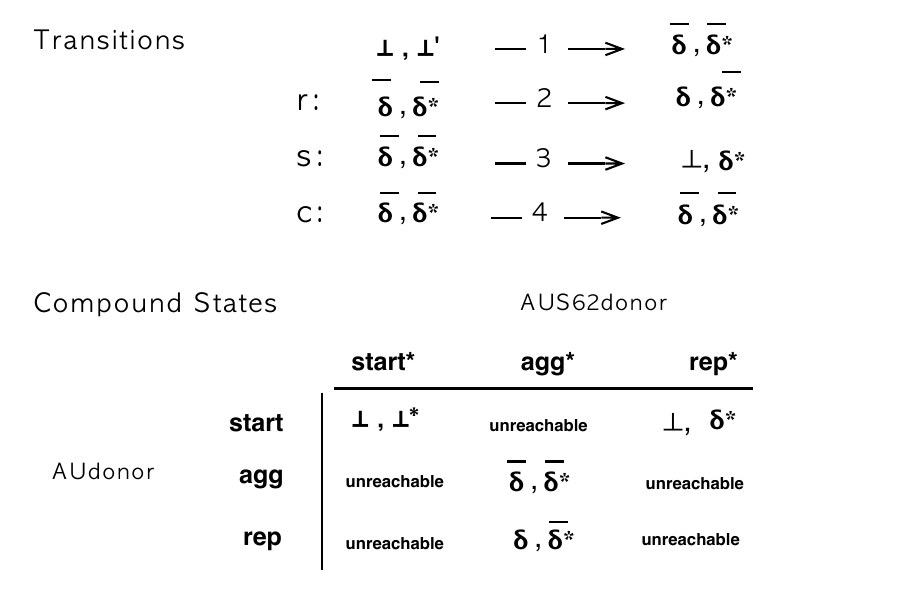}
\caption{\it Reachable Reporting States (Q1) (NB accumulation is asymmetric - All donations count towards S62 threshold (agg*), but only non-S62 donations count towards AUdonor (agg))}
\label{v}
\end{figure}

\begin{figure}[ht!]
\centering
\leavevmode
\includegraphics[width=0.5\textwidth,origin=c]{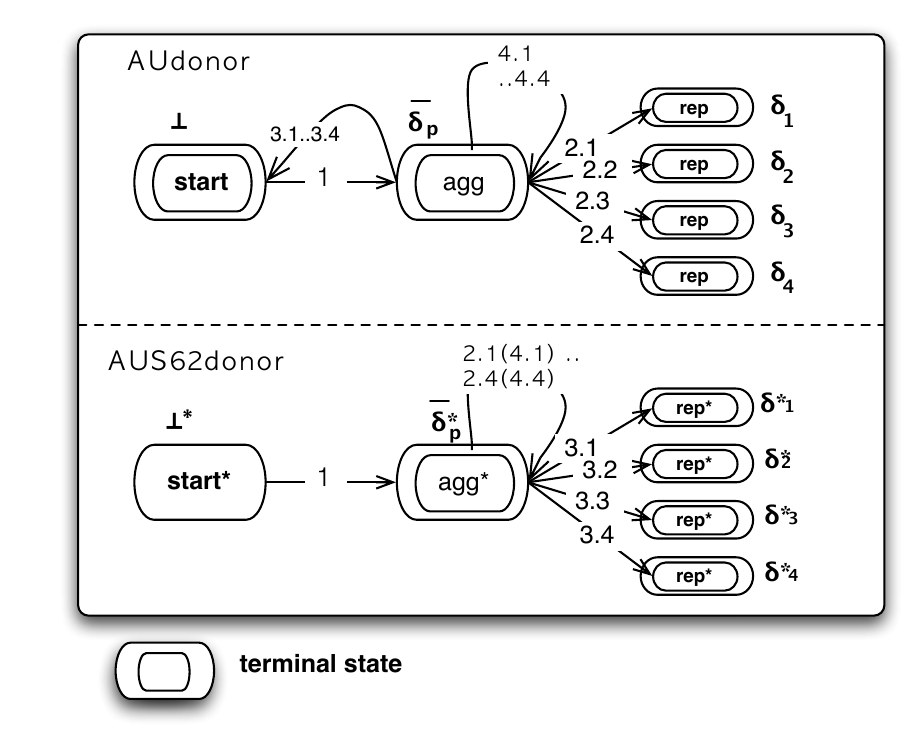}
\caption{\it Individual Donation Reporting State Chart (Q4)}
\label{vi}
\end{figure}

With these predicates in place we can revisit the act-decide transition system and prove that each formula that labels a state in the system is satisfied by an execution of the decision agent with appropriate data. We represent decision making as a state chart  (Figure~\ref{iv}~to~\ref{vi}) that accepts donations in the compound state {$ (\bot, \bot^*) $} and moves them to their external reporting states (if any) according to the following rules: 

\[ \delta_p \eqdef ( \delta(au_d,p) , \overline \delta^*(au_d,p) )
\]

\[ \delta^*_p \eqdef ( \bot, \delta^*(au_d,p)) 
\]

In  section~\ref{subsect:cover} we show that the test scenarios systematically explore just the permitted paths, as they are exercised over a year by two physical AU interacting through the virtual accounting unit for the donor.

The state chart  Figure~\ref{iv} shows how the classification proceeds in the first quarterly decision protocol. Here two concurrent state machines - AUdonor and AUS62donor - await a new set of donations from the donor to the AU for the period in a compound start state marked {$ (\bot, \bot^*) $}. When the period is closed all accepted donations for the unit in the year to date advance via transition (1) into the compound state {$ \overline {\delta }, \overline {\delta^*} $}. 

In this {$ \bf aggregation $} state each machine tests to see if the threshold is exceeded for the set of donations that have collected there. We distinguish the two distinct aggregation rules, that is, to include or exclude previously reported S62 donations, by the state labels {$ \bf agg $} (exclude) and {$ \bf agg* $} (include) (Figure~\ref{v}). 

Depending on the result of these tests, a set of AU donations will either transition (4) back to this compound state (and be recorded as CF), advance to a Quarterly Report (transition 2) or be reported under S62 (transition 3). 

Donation sets that advance to the S62 report in the AUS62donor machine will simultaneously revert to the AUdonor start state. This will exclude them from future AU donor threshold tests. 

At the end of this classification process the set of AU donations may be in any reachable compound state except {$ (\bot, \bot^*) $} (Figure~\ref{v}).

\textit{Each period this calculation will be repeated and all prior reports will be recalculated. This is necessary to revert backdated donations to the period in which their accepted date fell}. 

Figure~\ref{vi} shows the decision protocol AUdonor at the end of the fourth quarter. Donations which are not carried forward at the end of the year may have taken transition {$ 2.1, \ldots , 2.4 $} to be allocated to a Quarterly Report  {$ \delta_1, \ldots ,\delta_4 $}, or they may have taken transition {$ 3.1, \ldots ,3.4 $} to be included in aggregate in a Section 62(12) Report for the donor for a quarter {$ \delta^*_1, \ldots ,\delta^*_4 $}.

\subsection{Test Scenarios}
\label{subsect:scenarios}

Each quarter takes into account the history of donation reporting by the donor to the accounting unit. Depending whether Section 62(12) is active, or the AU has a prior Quarterly report from a donor, a recordable donation may either be carried forward, aggregrated in S62, or published in the Quarterly Report. We have shown how for each accounting unit and donor there are 15 possible histories, or AUdonor paths, linking reachable states of the reporting space (Figure~\ref{iiii}). In this section we consider how many combinations of AUdonor paths we need to prove that the reporting logic always works.

To do the proof we will need an additional convention to account for null reports. These are quarters when no donations are received by an accounting unit from a specific donor. Our \textbf {null reporting} convention [$ \bf{Rule 1b}$] mandates that if S62(12) is active in a quarter and the AU is not otherwise reporting, then where no donation is received by an AU from the donor, the S62 report will contain a notional null report. However, if the AU is reporting, that is the threshold for the AU has been breached in a prior quarter, then the notional donation will form part of the Quarterly report for the new quarter. If neither S62 or prior Quarterly reporting applies, then the null report is a notional Carried Forward entry.

We define a test \textbf{scenario} as a pair of permissible AUdonor test vectors, one for a local unit and the other for Head Office.  A test \textbf {vector} is a sequence of four (possibly null) donations that will signal some test path. Since we have 15 permitted AUdonor paths this means that there are potentially 15*15, that is 225, combinations of paths for each of which we need to construct a test scenario. However periods in which Section 62(12) is active in one AU cannot coexist with carried forward in another. We show in Lemma 1 that this reduces the number of combinations. Furthermore, as AU donor thresholds are constant values, we can set them all to the same value, without any loss of generality in our proof. We can therefore show, in Lemma 2, that by symmetry we only need consider each combination of paths once (most occur twice in the 225).

\begin{figure}[ht!]
\includegraphics[width=0.5\textwidth]{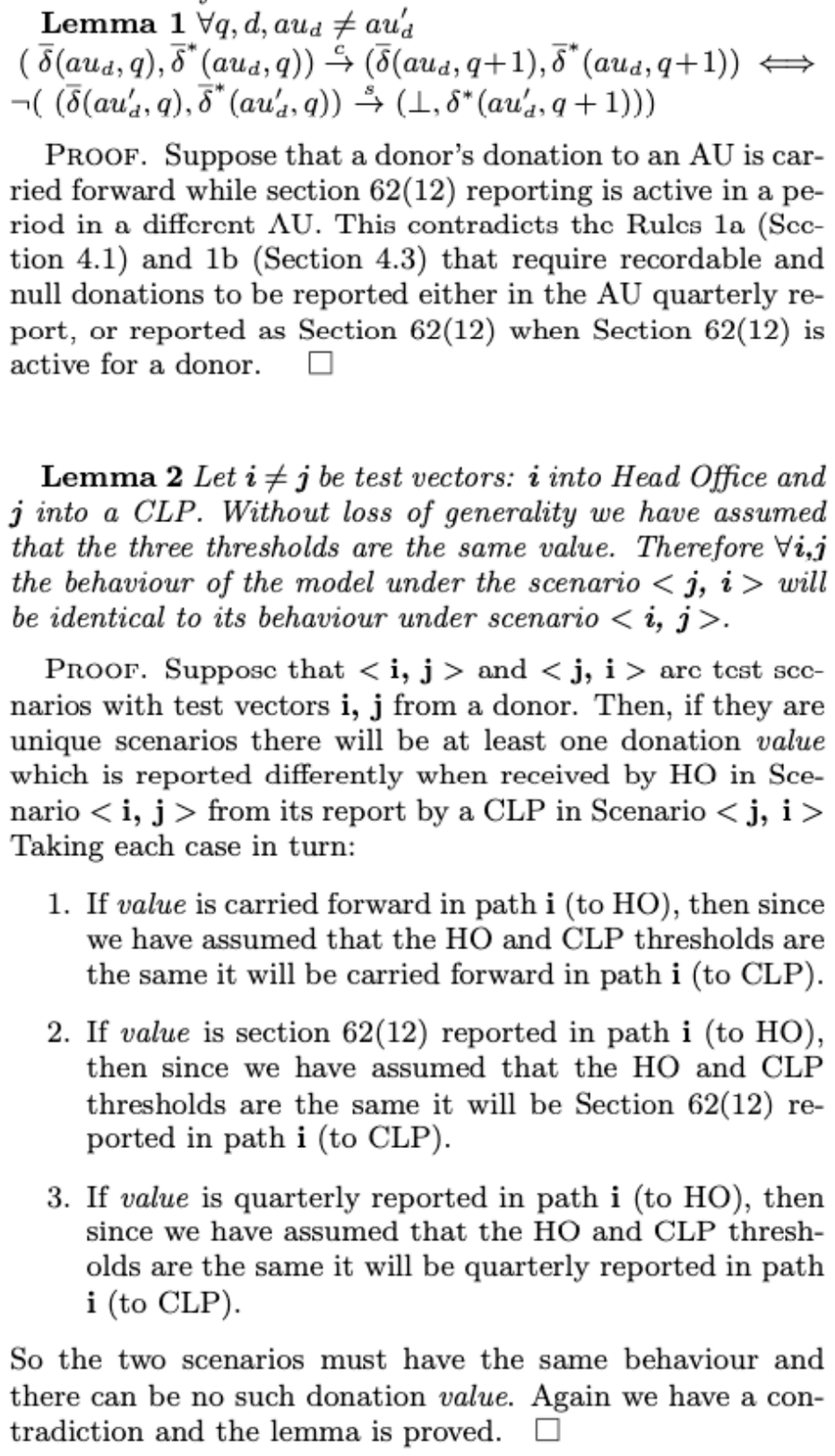}
\caption{\it Lemmas}
\label{lemmas}
\end{figure}

The figure~\ref{lemmas} gives two lemmas we will need and their proofs.

\begin{figure}[ht!]
\includegraphics[width=0.5\textwidth]{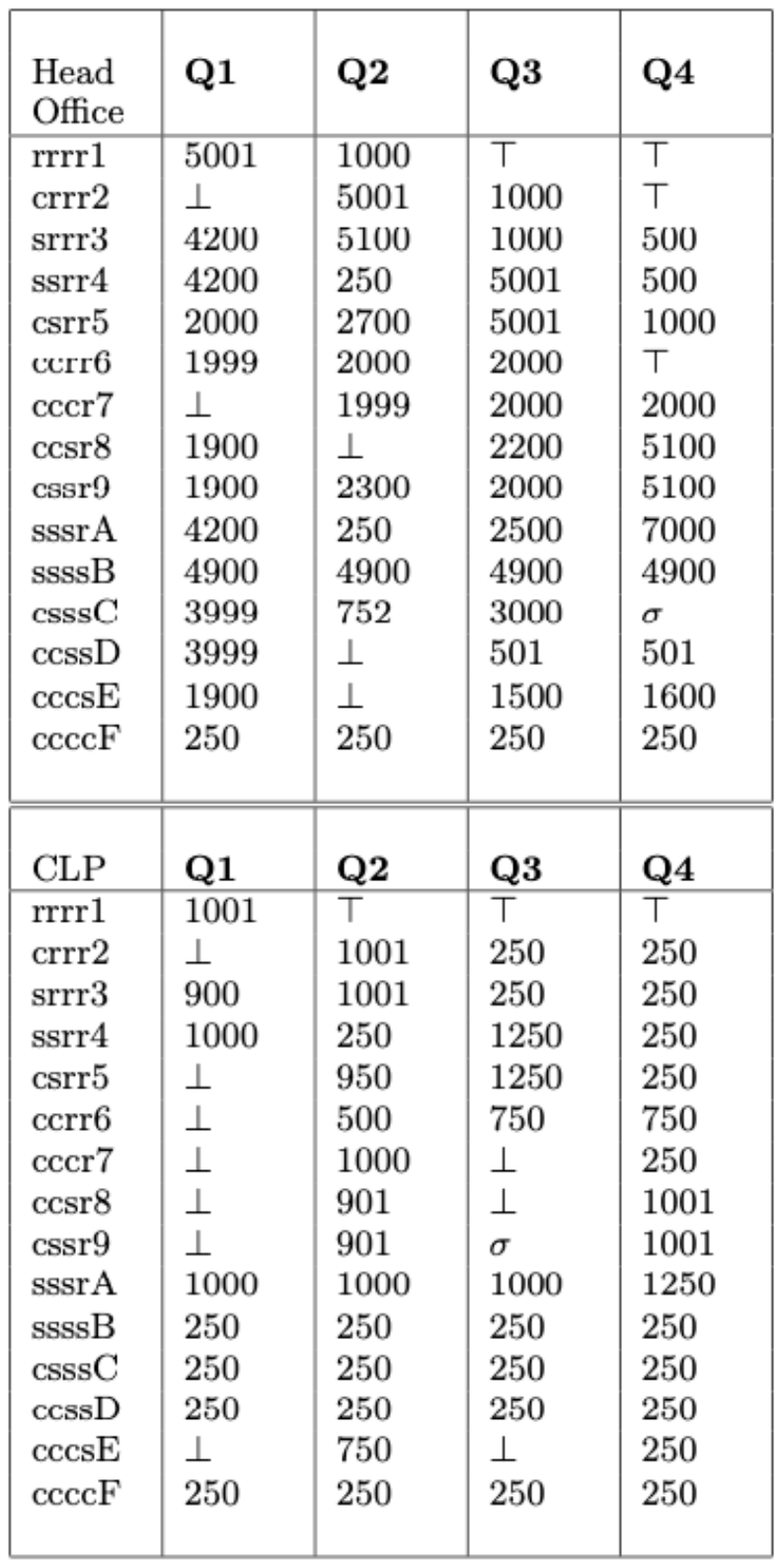}
\caption{\em Vectors signaling Paths to HO and to a CLP}
\label{objectives:allvectors}
\end{figure}

\subsection{Coverage Proof}

%
%

\begin{figure}[ht!]
\includegraphics[width=0.5\textwidth]{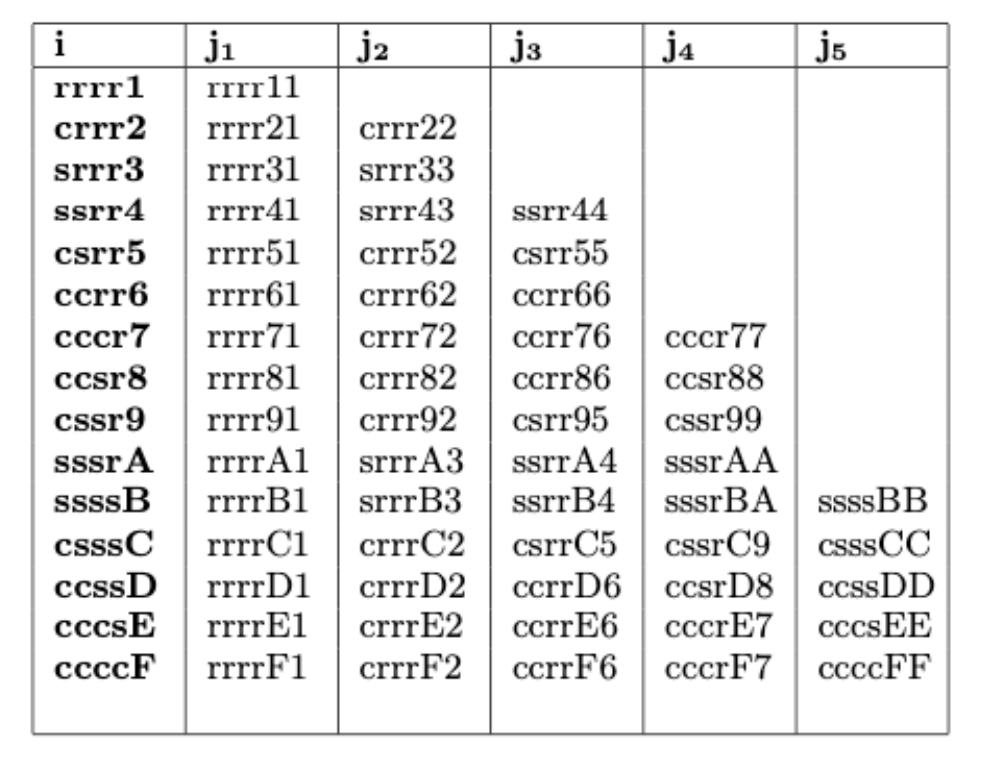}
\caption{\em AUdonor Test scenarios: permitted j where i is fixed}
\label{objectives:allscenes}
\end{figure}

\label{subsect:cover}

Using these lemmas we can create an exhaustive set of test scenarios to encompass all the permissible combinations. 

\begin{proof}

In Figure~\ref{objectives:allscenes} each unique test scenario {$ <\textbf{i, j}> $} is represented by a label hexadecimal ``ij''  {$ i \ge j $} where ``i,j''  are hexadecimal path numbers (Figure~\ref{objectives:summary})
and {$ i \ge j $}. By inspection there are no other pairs that are permitted by Lemma 1.

Test vectors are shown in Figure~\ref{objectives:allvectors}. 

This shows one test vector for each path $ \pi $ to HO and to a CLP: we can call them HO($ \pi $) and CLP($ \pi $). 

We use {$ \bot $} to indicate that a null carried forward donation is reported, {$ \top $} to indicate a null donation in a Quarterly Report, and {$ \sigma $} for a null donation in a Section 62 (12) Report (Rule 1b Section~\ref{subsect:scenarios}).

From Lemma 2 it follows that for every  test scenario  {$ <~HO(\pi_i), CLP(\pi_j)>  $}  in table~\ref{objectives:allscenes} there is a test scenario of the form {$ <HO(\pi_j),CLP(\pi_i)> $}.  

By construction there are no other test scenarios. \qed 


\end{proof}

\section{Conclusion}
\label{subsect:conc}

In this paper we have developed a new computational theory to account for voluntary behaviour for the common good. We built on a vocabulary of objects, operations and relationships first proposed by the economist Mancur Olson in his ``Logic of Collective Action.'' Olson predicted that action for the common good would only be taken voluntarily in small groups where the total cost of the first production of the good could be met profitably by one person. In all other cases coercion would be needed.

We tested his predictions by observing first the promotion, then the rejection and finally the compliance of the UK Labour Party with the Political Parties, Elections and Referendums' Act (2000). We found that compliance was achieved after two annual reporting cycles without augmenting the sanctions regime -- which would have been needed to deal with the initial non-compliance in Olson's theory. Rather compliance was achieved by reifying the process of inner and intra-unit negotiation. The process model was used to win confidence in a recording and reporting system built to this specification by means of a formal proof of test coverage. We found that confidence -- a key aspect of functioning teams, markets and systems -- was a category of common good that fell outside Olson's theory.

The language used in the Act itself to describe the operation of PPERA is much more complex than our structured mathematical treatment. This lends itself to intuitive diagram conventions such as those used in a Conversation for Action or London Underground map and in our act-decide transition system. 

\subsection{Directions for Research}

Further research is needed on techniques to explore the application of this approach in other public services, such as health, welfare or education. In particular new tools and methods are needed to automate test coverage analysis for decision protocols. 

There remain major risks of system development and process failure in the case of large scale distributed system design. One of the most egregious of these has occurred in the UK where a Fujitsu designed distributed system led to the biggest ever miscarriage of justice.  700 sub-postmasters were wrongfully accused of theft. The trial judge accused Post Office executives and computer operatives of covering up the truth and even lying  \cite{b13}.

We have shown that the root cause of these failures may lie not only in the limitations of traditional systems analysis, first highlighted by Winograd and Flores in ``Computers and Cognition'' but also in archaic forms of parliamentary draughtsmanship, outmoded economic thought and system development methods that are no longer fit for purpose.

\section{Acknowledgments}

The author is grateful to Mikula\v{s} Teich, Kristen Nygaard, Charles Clarke, Paul Corrigan, Terry Winograd, Claudio Ciborra, David Holloway, William Clocksin, Ben Galewsky and others for discussions over many years on the successes and failures of systems development and to the anonymous reviewers of earlier drafts of this paper which led to improvements.

\bibliographystyle{\style}

\section*{Appendix: Section 62(12)}
\label{sec:six}
Clause 12 of Section 62 of the PPERA, 2000 sets out the report to be made on donations that need to be reported by virtue of their forming part of an aggregate donation from a counterparty whose elements - taken by themselves - would not otherwise warrant reporting. 

The form of the quarterly report to be made is set out in \textit{Schedule 6}.

\subsection*{PPERA, Section 62}

The clauses of Section 62 are as follows. 

\label{sec:621}
(1) The treasurer of a registered party shall, in the case of
each year, prepare a report under this subsection in respect of each
of the following periods-

(a) January to March;

(b) April to June;

(c) July to September;

(d) October to December.

(2) In this section-
"donation report" means a report prepared under subsection (1);

"reporting period", in relation to such a report, means the
period mentioned in any of paragraphs (a) to (d) of that
subsection to which the report relates.

\label{sec:623}
(3) The donation reports for any year shall, in the case of each
permissible donor from whom any donation is accepted by the
party during that year, comply with the following provisions of this
section so far as they require any such donation to be recorded in a
donation report; 

and in those provisions any such donation is referred to, in relation to the donor and that year, as a "relevant donation".

(4) Where no previous relevant donation or donations has or
have been required to be recorded under this subsection, a relevant
donation must be recorded-

(a) if it is a donation of more than £5,000, or

(b) if, when it is added to any other relevant donation or
donations, the aggregate amount of the donations is more
than £5,000.

(5) A donation to which subsection (4) applies must-

(a) (if within paragraph (a) of that subsection) be recorded in
the donation report for the reporting period in which it is
accepted, or

(b) (if within paragraph (b) of that subsection) be recorded
(as part of the aggregate amount mentioned in that
paragraph) in the donation report for the reporting period in
which the donation which causes that aggregate amount to
be more than £5,000 is accepted.

(6) Where any previous relevant donation or donations has or
have been required to be recorded under subsection (4), a relevant
donation must be recorded at the point when there has or have been
accepted-

(a) since the donation or donations required to be recorded
under subsection (4), or

(b) if any relevant donation or donations has or have
previously been required to be recorded under this
subsection, since the donation or donations last required to
be so recorded, any relevant donation or donations of an amount or aggregate
amount which is more than £1,000.

(7) A donation to which subsection (6) applies on any occasion
must-

(a) if it is the only donation required to be recorded on that
occasion, be recorded in the donation report for the reporting
period in which it is accepted, or

(b) in any other case be recorded (as part of the aggregate
amount mentioned in that subsection) in the donation report
for the reporting period in which the donation which causes
that aggregate amount to be more than £1,000 is accepted.

(8) For the purposes of subsections (4) to (7) as they apply in
relation to any year-

(a) each payment to which section 55(2) applies and which is
accepted by the party during that year shall be treated as a
relevant donation in relation to that year, and

(b) each payment to which section 55(3) applies and which
is received from a particular donor and accepted by the party
during that year shall be treated as a relevant donation in
relation to the donor and that year;

and the donation reports for the year shall accordingly comply with
subsections (4) to (7) so far as they operate, by virtue of paragraph
(a) or (b) above, to require any relevant donation falling within that
paragraph to be recorded in a donation report.

(9) A donation report must also record every donation falling
within section 54(1)(a) or (b) and dealt with during the reporting
period in accordance with section 56(2).

(10) If during any reporting period-
(a) no donations have been accepted by the party which, by
virtue of the preceding provisions of this section, are
required to be recorded in the donation report for that period, and 

(b) no donations have been dealt with as mentioned in
subsection (9), the report shall contain a statement to that effect.

(11) Where a registered party is a party with accounting units,
subsections (3) to (10) shall apply separately in relation to the
central organisation of the party and each of its accounting units-

(a) as if any reference to the party were a reference to the
central organisation or (as the case may be) to such an
accounting unit; but

(b) with the substitution, in relation to such an accounting
unit, of "£1,000" for "£5,000" in each place where it occurs
in subsections (4) and (5).

(12) However, for the purposes of subsections (3) to (7) in their
application in relation to the central organisation and any year by
virtue of subsection (11), any donation-
(a) which is accepted from a permissible donor by any of the
accounting units during that year, but
(b) which is not required to be recorded under subsection (4)
or (6) (as they apply by virtue of subsection (11)) as a
donation accepted by the accounting unit, shall be treated as a donation accepted from the donor during that year by the central organisation.

(13) Schedule 6 has effect with respect to the information to be
given in donation reports.

\subsection*{Schedule 6} \label{Sec:subsix}

1. - (1) In this Schedule-
(a) "quarterly report" means a report required to be prepared by virtue of section 62;
(b) "weekly report" means a report required to be prepared by virtue of section 63; and 
"recordable donation", in relation to a quarterly or weekly
report, means a donation required to be recorded in that report.

(2) References in this Schedule to a registered party shall, in thecase of a party with accounting units, be read as references to the central organisation of the party.

Identity of donors: quarterly reports

2. - (1) In relation to each recordable donation (other than one to which paragraph 6 or 7 applies) a quarterly report must give the following information about the donor -

(a) the information required by any of sub-paragraphs (2) to
(10), (12) and (13) below; and 

(b) such other information as may be required by regulations made by the Secretary of State after consulting the Commission; or, in the case of a donation falling within sub-paragraph (11) below, the information required by that sub-paragraph.

(2) In the case of an individual the report must give his full name
and-
(a) if his address is, at the date of receipt of the donation, shown in an electoral register (within the meaning of section 54), that address; and
(b) otherwise, his home address (whether in the United Kingdom or elsewhere).

(3) Sub-paragraph (2) does not apply in the case of a donation in the form of a bequest, and in such a case the report must state that the donation was received in pursuance of a bequest and give-

(a) the full name of the person who made the bequest; and

(b) his address at the time of his death or, if he was not then
registered in an electoral register (within the meaning of section 54) at that address, the last address at which he was so registered during the period of five years ending with the date of his death.

(4) In the case of a company falling within section 54(2)(b) the report must give-

(a) the company's registered name;

(b) the address of its registered office; and

(c) the number with which it is registered.

(5) In the case of a registered party the report must give-

(a) the party's registered name; and

(b) the address of its registered headquarters.

(6) In the case of a trade union falling within section 54(2)(d) the
report must give-

(a) the name of the union, and

(b) the address of its head or main office, as shown in the list kept under the Trade Union and Labour

(7) In the case of a building society within the meaning of the
Building Societies Act 1986, the report must give-

(a) the name of the society; and

(b) the address of its principal office.

(8) In the case of a limited liability partnership falling within
section 54(2)(f), the report must give-

(a) the partnership's registered name; and

(b) the address of its registered office.

(9) In the case of a friendly or other registered society falling
within section 54(2)(g) the report must give-

(a) the name of the society, and

(b) the address of its registered office.

(10) In the case of an unincorporated association falling within
section 54(2)(h) the report must give-

(a) the name of the association; and

(b) the address of its main office in the United Kingdom.

(11) In the case of a payment to which section 55(2) applies the
report must give the statutory or other provision under which it was
made.

(12) In the case of a donation to which section 55(3) applies, the
report must give the full name and address of the donor.

(13) In the case of a donation to which section 55(5) applies, the
report must state that the donation was received from a trustee,
and- 
(a) in the case of a donation falling within section 162(2), give-

(i) the date on which the trust was created, and

(ii) the full name of the person who created the trust and of every other person by whom, or under whose will, property was transferred to the trust before 27th July 1999, and

(b) in the case of a donation falling within section 162(3),
give in respect of-

(i) the person who created the trust, and,

(ii) every other person by whom, or under whose will,
property has been transferred to the trust, the information which is required by any of sub-paragraphs (2) to (10) to be given in respect of the donor of a recordable
donation.

(14) In this Act or the Representation of the People Act 1983
any reference (however expressed) to information about the donor
of a donation which is framed by reference to this paragraph is, in
relation to such a donation as is mentioned in paragraph (a) or (b)
of sub-paragraph (13), a reference to information about every
person specified in paragraph (a) or (b) of that sub-paragraph.
Identity of donors: weekly reports

3. In relation to each recordable donation a weekly report must
give all such details of the name and address of the donor as are for
the time being known to the party.

Value of donation

4. - (1) In relation to each recordable donation a quarterly or
weekly report must give the following details about the donation.

(2) If the donation was a donation of money (in cash or
otherwise) the report must give the amount of the donation.

(3) Otherwise the report must give details of the nature of the
donation and its value as determined in accordance with section 53.

Circumstances in which donation made

5. - (1) In relation to each recordable donation a quarterly or
weekly report must-

(a) give the relevant date for the donation; and

(b) (in the case of a quarterly report)-

(i) state whether the donation was made to the
registered party or any accounting unit of the party; or

(ii) in the case of a donation to which section 62(12)
applies, indicate that it is a donation which falls to be
treated as made to the party by virtue of that provision.

(2) In the case of a donation to which section 55(3) applies, the
report must in addition give-

(a) the date or dates on or between which the visit to which
the donation relates took place, and

(b) the destination and purpose of the visit.

(3) For the purposes of this paragraph as it applies to a quarterly
report, the relevant date for a donation is-

(a) (if within section 62(4)(a) or (7)(a)) the date when the
donation was accepted by the party or the accounting unit;

(b) (if within section 62(4)(b) or (7)(b)) the date when the
donation was accepted by the party or the accounting unit
which caused the aggregate amount in question to be more
than the limit specified in that provision;

(c) (if within section 62(9)) the date when the donation was
received.

(4) For the purposes of this paragraph as it applies to a weekly
report, the relevant date for a donation is the date when the
donation was received by the party or its central organisation as
mentioned in section 63(3).

Donations from impermissible donors

6. In relation to each recordable donation to which section
54(1)(a) applies a quarterly report must-
(a) give the name and address of the donor; and
(b) give the date when, and the manner in which, the
donation was dealt with in accordance with section 56(2)(a).
Donations from unidentifiable donors

7. In relation to each recordable donation to which section
54(1)(b) applies a quarterly report must give-
(a) details of the manner in which it was made,
(b) details of any element of deception or concealment
employed by the donor of which the registered party or any
accounting unit of the party became aware and the means by
which it was revealed; and
(c) the date when, and the manner in which, the donation
was dealt with in accordance with section 56(2)(b).

Other details

8. A quarterly or weekly report must give such other information
(if any) as is required by regulations made by the Commission.
Political Parties, Elections and Referendums Act 2000


\end{document}